\documentclass[aps,prl,twocolumn,showpacs,superscriptaddress]{revtex4-2}
\usepackage[T1]{fontenc}
\usepackage{amsmath,amssymb}
\usepackage{graphicx}
\usepackage{subfigure}
\usepackage{verbatim}
\usepackage{amsfonts}
\usepackage{dcolumn}
\usepackage{bm}
\usepackage{ulem}
\usepackage{color}
\usepackage[colorlinks,citecolor=blue]{hyperref}

\usepackage{mathrsfs} 
\begin{document}

\title{Hierarchical Trion Formation and Fractionalized  Solitons in One Dimension}

\author{Yan-Guang Yue}
\affiliation{Department of Physics and State Key Laboratory of Surface Physics, Fudan University, Shanghai 200433,  China}

\author{Qi Song}
\affiliation{Department of Physics and State Key Laboratory of Surface Physics, Fudan University, Shanghai 200433,  China}

\author{Jie Lou}
\affiliation{Department of Physics and State Key Laboratory of Surface Physics, Fudan University, Shanghai 200433,  China}

\author{Yan Chen}
\email{yanchen99@fudan.edu.cn}
\affiliation{Department of Physics and State Key Laboratory of Surface Physics, Fudan University, Shanghai 200433,  China}
\affiliation{Shanghai Branch, Hefei National Laboratory, Shanghai 201315,  China}

\begin{abstract}
Stabilizing commensurate $2:1$ trions in one-dimensional quantum mixtures is typically hindered by phase separation. Within the conventional density-driven framework, the asymmetric locking mechanism generically couples to a softening density mode, creating a geometric constraint that suppresses the formation of a stable trionic liquid.
 We demonstrate that correlated kinetics can reorganize the low-energy structure to bypass this instability. Driven by microscopic fermion pair-hopping, the fermions first form an emergent pair liquid. At a $2:1$ filling, the density of the emergent pairs exactly matches the bosons. The low-energy theory is therefore reorganized into an effective symmetric $1:1$ pair-boson mixture. This symmetry decouples the locking mechanism from the softening mode, preempting phase separation and stabilizing a trionic liquid. The reconstructed phase exhibits a correlation hierarchy where only the composite trion retains quasi-long-range order, while the gapped relative sector supports parity-constrained topological kink excitations.
\end{abstract}

\maketitle

\textit{Introduction.}---
The realization of quantum states of matter in multicomponent ultracold atomic gases is a central focus in strongly correlated physics. While two-body phenomena such as Cooper pairing and diatomic molecule formation are well established \cite{Bloch2008, chin2010feshbach}, the synthesis of higher-order multi-particle bound states---such as trions and tetramers---remains a captivating challenge \cite{dalmonte2011cluster, petrov2004three,Roux2011,Kagan2004, Huan2022, Duda2023}. One-dimensional (1D) Bose-Fermi mixtures, characterized by enhanced quantum fluctuations and highly tunable interactions, provide a pristine platform for exploring these composite liquids \cite{cazalilla2011one, goral2002quantum,CazalillaHo2003,Decamp_2017,Gunter2006, Ospelkaus2006}. 

Despite intense experimental and theoretical interest, stabilizing an asymmetric composite phase, such as a commensurate $2:1$ trionic liquid, poses a distinct physical challenge . In the standard density-driven framework, composite formation relies on interspecies attraction. While previous studies have demonstrated that $2:1$ trionic locking can be stabilized by engineering strong mass or velocity asymmetries to suppress competing fluctuations \cite{Roux2011, RizziImambekov2007}, the natural symmetric-velocity regime is generically unstable. In this regime, the bare asymmetric locking mechanism necessarily couples to the softening total-density mode. As a result, phase separation dominates before trionic locking can fully develop. Consequently, phase separation typically dominates the low-energy physics \cite{CazalillaHo2003, Pollet2006}, preempting the establishment of trionic coherence.

To stabilize the trionic phase without relying on asymmetric parameter tuning, we investigate how correlated kinetics can reorganize the system. We demonstrate that introducing a microscopic fermion pair-hopping process \cite{gotta2021prl, gotta2021prb,Gotta2022} provides a direct route to composite formation. Large pair hopping suppresses low-energy single-fermion excitations, causing the fermions to first form a paired liquid. Because the macroscopic filling is $\rho_c : \rho_b = 2 : 1$, the density of the emergent pairs exactly matches the bosons ($\rho_P = \rho_b$). This density matching converts the original asymmetric $2:1$ locking channel into an effective symmetric $1:1$ pair-boson resonance, allowing the system to safely bypass the phase-separation instability and stabilize a trionic liquid.

The resulting composite phase is characterized by a distinct mode structure: the effective locking mechanism gaps out the relative-density fluctuations while preserving a gapless total-density acoustic mode. This establishes a strict correlation hierarchy. We show that single-particle and partial composite correlations decay exponentially, leaving the full $2:1$ trion as the unique quasi-long-range ordered channel. Furthermore, the gapped relative sector supports fractionalized geometric kinks. Because these phase slips change fermion parity, they are confined within the even-parity pair liquid, leaving the composite $2\pi$ instanton as the true local topological defect. These results establish the trionic liquid as a stable composite phase
with a gapless acoustic mode,
confined partial excitations,
and parity-constrained topological defects,
providing a theoretical framework that complements previous numerical studies \cite{song2025numerical}
of correlated Bose–Fermi mixtures.


\textit{Model and Physical Motivation.}---We consider a 1D mixture of spinless fermions and hard-core bosons governed by the extended Bose-Fermi Hubbard Hamiltonian $H = H_0 + H_{\text{int}} + H_{cc}$. The standard single-particle kinetic and interspecies interaction terms are
\begin{equation}
H_0 + H_{\text{int}} = -\sum_{j, \nu \in \{c,b\}} t_\nu \left( \nu_j^\dagger \nu_{j+1} + \text{h.c.} \right)+U_{bc} \sum_j n_{c,j} n_{b,j},
\end{equation}
where $c_j$ and $b_j$ are the microscopic annihilation operators, $t_\nu$ are the single-particle hopping amplitudes, and $U_{bc}< 0$ represents the interspecies density-density attraction. We focus on the macroscopic commensurate filling ratio $\rho_c : \rho_b = 2 : 1$, which provides the necessary particle reservoir for forming $2:1$ trionic composites. To dynamically reorganize the low-energy physics, the model includes a correlated fermion pair-hopping term $H_{cc} = -t_{cc} \sum_j ( c_{j-1}^\dagger c_j^\dagger c_j c_{j+1} + \text{h.c.} )$.

\begin{figure}[t]
    \centering
    \includegraphics[width=\columnwidth]{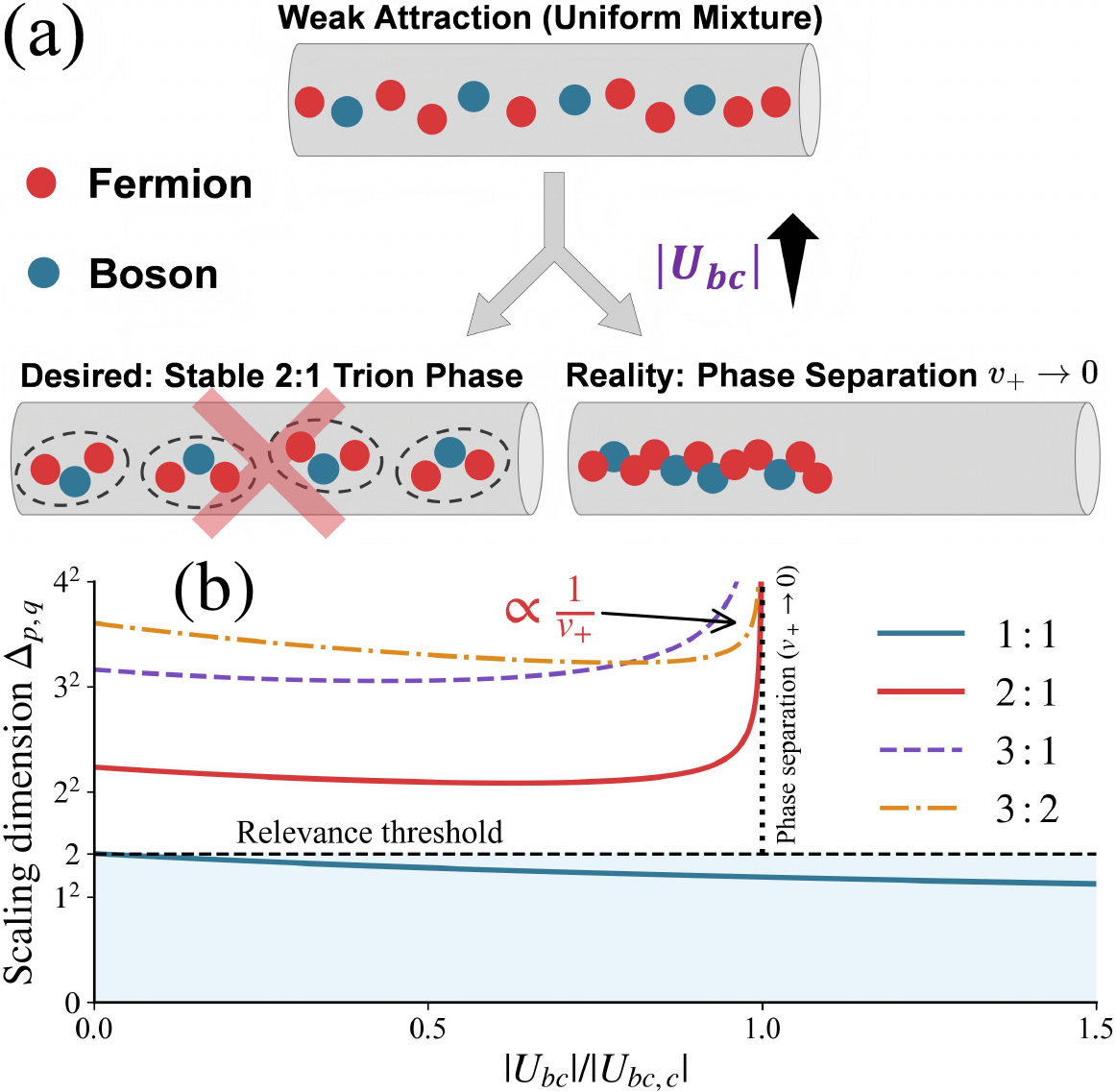}
    \caption{
    (a) Physical origin of the instability of the bare $2:1$ trionic mechanism. Increasing attraction softens the compressible density mode and eventually drives phase separation before stable trionic locking can develop.
    (b) Scaling dimensions of several commensurate locking channels near the phase-separation boundary. As the density mode softens upon approaching phase separation, the scaling dimensions of the asymmetric locking operators rapidly increase, rendering the locking mechanism irrelevant.
 In contrast, the symmetric $1:1$ channel remains free from this divergence. The analytical derivation of the scaling dimensions is provided in the Supplemental Material.
    }
    \label{fig:bare_instability}
\end{figure}

To understand why the pair-hopping term becomes essential for stabilizing the trionic phase, it is instructive to first examine the generic instability of the bare framework ($t_{cc} = 0$), as illustrated in Fig.~\ref{fig:bare_instability}.
In the low-energy continuum limit, Abelian bosonization \cite{Giamarchi2003} describes the density fluctuations via the fields $\phi_c$ and $\phi_b$. At the commensurate filling $\rho_c = 2\rho_b$, the attractive interaction generates a density-locking term of the form
$
H_{\text{locking}} \propto \int dx \cos(4\phi_b - 2\phi_c),
$
which favors the formation of a $2:1$ trionic state.

For the system to establish composite coherence, this locking term must dominate the low-energy physics. However, increasing the interspecies attraction $U_{bc}$ to strengthen this locking simultaneously increases the overall compressibility of the mixture. This causes the total-density acoustic mode to strongly soften, driving the system toward mechanical instability and phase separation.

Crucially, because the trion contains unequal particle numbers, the asymmetric $2:1$ locking term necessarily mixes with this compressible density mode. In the natural symmetric-velocity regime ($v_c \approx v_b$), this inherent coupling strictly ties the fate of the trionic resonance to the mechanical instability of the mixture. Consequently, the softening of the compressible mode strongly suppresses the locking tendency. While extreme tuning of the mass or velocity asymmetry can suppress the coupling between the locking tendency and the soft compressible mode \cite{Roux2011}, in the symmetric regime, phase separation generally preempts the establishment of trionic coherence. 

This demonstrates that ordinary density-driven attraction is generally insufficient to stabilize the trionic liquid in the symmetric regime, motivating the introduction of correlated pair hopping as an alternative route to trionic stabilization. As we show below, large pair hopping suppresses low-energy single-fermion motion and reorganizes the infrared theory into an effective pair-boson liquid.

\begin{figure}[t]
\centering
\includegraphics[width=\columnwidth]{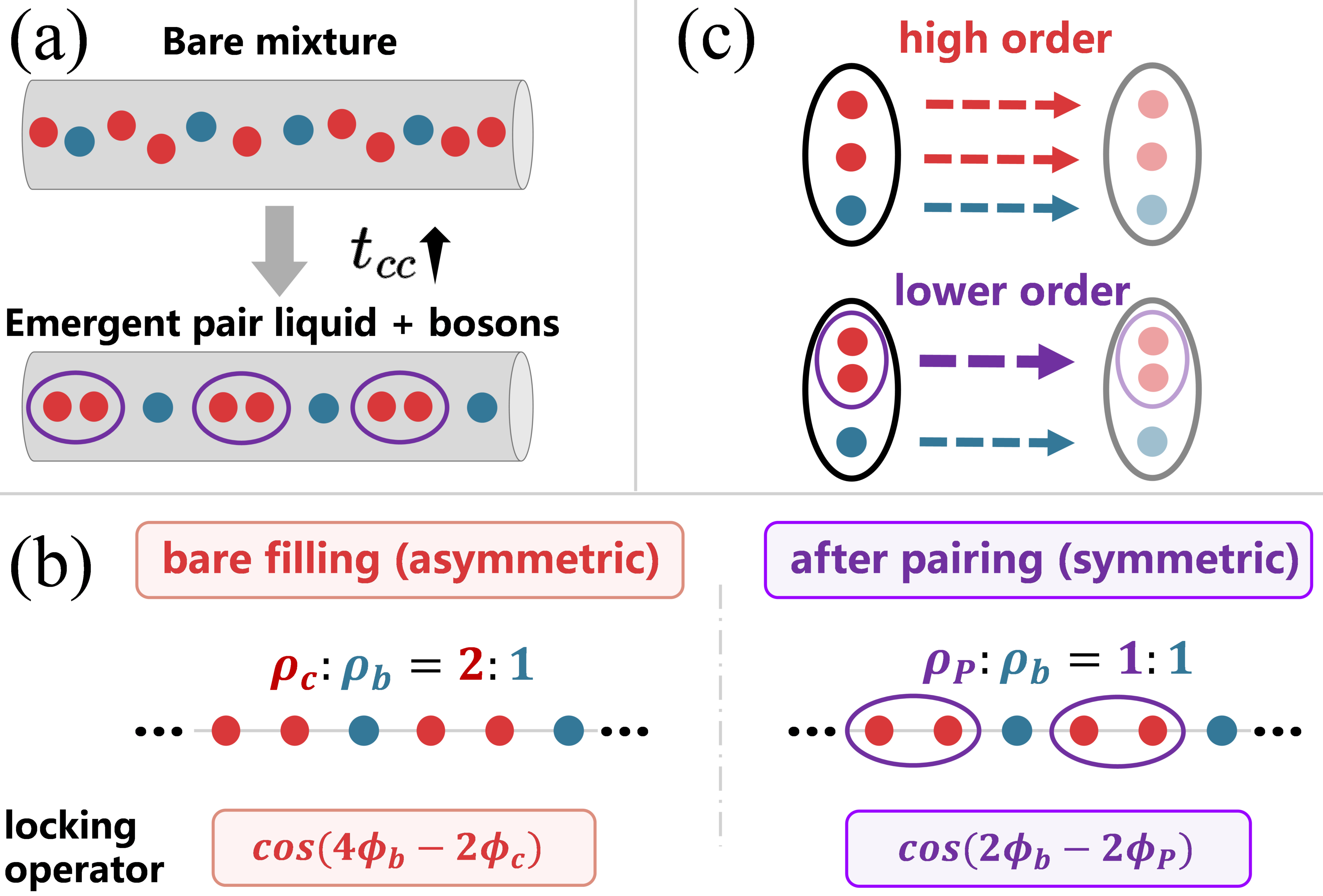}
\caption{Schematic illustration of the infrared reconstruction driven by correlated pair hopping.
(a) Large $t_{cc}$ suppresses isolated low-energy fermions and reorganizes the system into an emergent pair liquid coexisting with bosons.
(b) At $\rho_c:\rho_b=2:1$, the emergent pair density satisfies $\rho_P=\rho_b$, converting the original asymmetric locking term $\cos(4\phi_b-2\phi_c)$ into the symmetric pair-boson locking term $\cos(2\phi_b-2\phi_P)$.
(c) Microscopic illustration of the effective transport processes: without pair hopping, trionic motion arises only through higher-order virtual processes, whereas correlated pair hopping reduces the transport to a lower-order process.
}
\label{fig:fig2}
\end{figure}

\textit{Emergent Pair-Boson Description.}---We now introduce the correlated pair hopping $t_{cc}$ to overcome the instability of the bare model. Microscopically, large pair hopping suppresses isolated low-energy single-fermion excitations. As illustrated schematically in Fig.~\ref{fig:fig2}(a), the fermions form an emergent pair liquid. A Schrieffer--Wolff analysis \cite{SW, BRAVYI20112793} (see Supplemental Material) further shows that the pair hopping generates an effective correlated motion of a boson and a fermion pair. As illustrated in Fig.~\ref{fig:fig2}(c), this process reduces the transport from a high-order virtual process in the bare model to a lower-order process in the reconstructed pair-liquid background.

The low-energy fermionic sector is therefore dominated by an emergent pair liquid, described by a continuous pair density field $\phi_P$. Because the system is fixed at the commensurate macroscopic filling $\rho_c = 2\rho_b$, the density of these emergent pairs exactly matches the boson density:
$
\rho_P = \frac{1}{2}\rho_c = \rho_b.
$

This exact density matching converts the original asymmetric $2:1$ problem into an effective symmetric $1:1$ pair-boson mixture [Fig.~\ref{fig:fig2}(b)]. Consequently, the low-energy locking potential reduces to a symmetric form: the bare high-order resonance $\cos(4\phi_b - 2\phi_c)$ is replaced by an effective pair-boson locking term,
$H_{\text{eff}} \propto \int dx \cos(2\phi_b - 2\phi_P)$.

This mapping directly resolves the instability of the bare model. In the original asymmetric resonance $\cos(4\phi_b-2\phi_c)$, the unequal particle numbers of the two species generate a finite coupling to the softening total-density mode. In contrast, the effective pair-boson locking term $\cos(2\phi_b-2\phi_P)$ is perfectly symmetric, causing this coupling to vanish.

Thus, the effective locking term no longer couples to the softening compressible mode. The pair formation and subsequent density matching remove the geometric obstruction present in the bare theory. Physically, the pair hopping reorganizes the asymmetric three-body problem into a symmetric locking problem between bosons and emergent fermion pairs.


\textit{Stable Trionic Liquid and Mode Structure.}---Because the symmetric pair-boson locking operator is decoupled from the softening compressible mode, the system undergoes a Berezinskii-Kosterlitz-Thouless transition \cite{Christodoulou2021,Kosterlitz1973} into a locked trionic liquid before phase separation can occur. This transition establishes a distinct low-energy mode structure. 

The effective locking potential gaps out the relative-density sector, defined by the field $\Phi^*_- \equiv \phi_b - \phi_P$, pinning it near the minima of the locking potential. Concurrently, the total-density sector, defined by $\Phi^*_+ \equiv \phi_b + \phi_P$, completely decouples from the locking potential. This acoustic mode remains gapless, governing the collective superfluid density fluctuations of the composite liquid. The physical properties of the phase are therefore determined by the interplay of this massive relative sector and the gapless acoustic background.


\begin{figure}[t]
\includegraphics[width=\columnwidth]{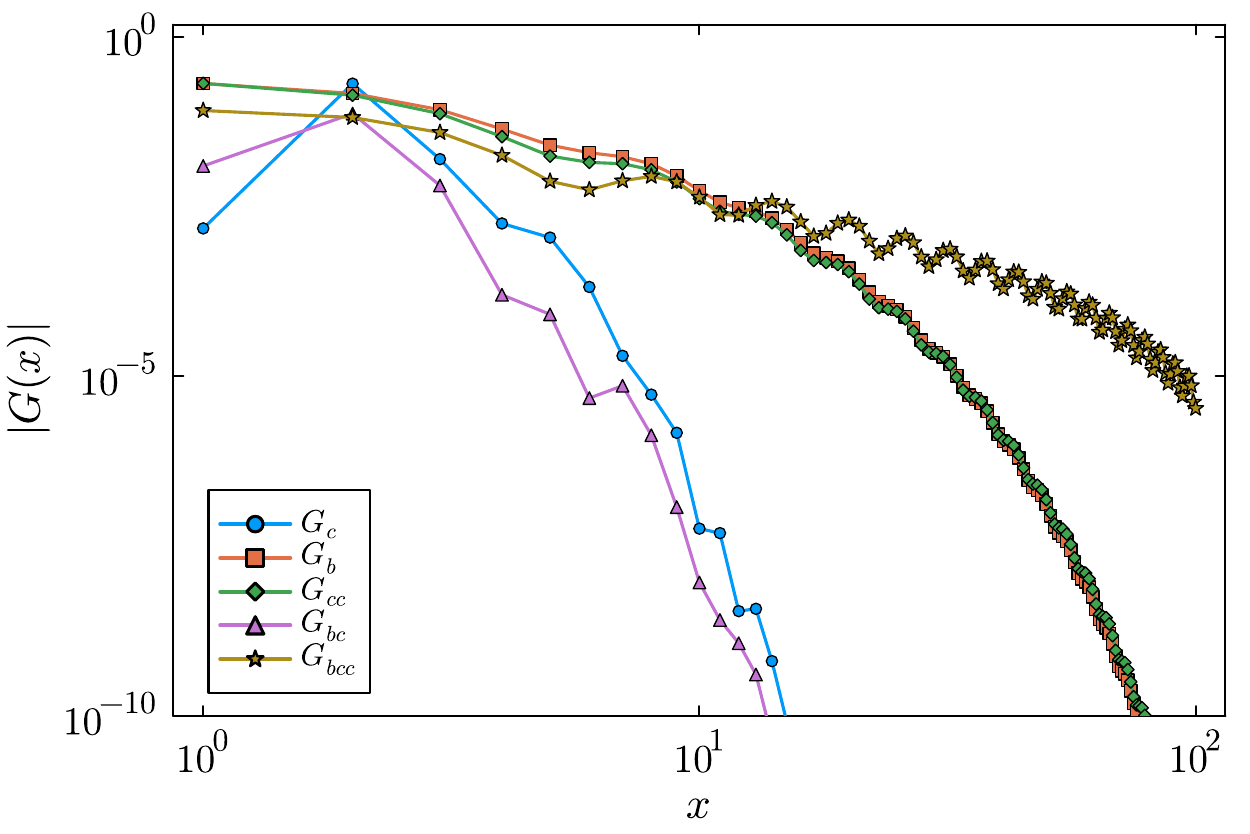}
\caption{Correlation hierarchy in the reconstructed trionic liquid.
Shown are the single-particle correlations
$G_c(x)$ and $G_b(x)$,
the partial composite correlations
$G_{cc}(x)$ and $G_{bc}(x)$,
and the full trionic correlation
$G_{bcc}(x)$.
The single-particle and partial composite channels decay rapidly,
whereas the trionic correlation remains the dominant long-distance channel,
displaying algebraic decay.
This behavior demonstrates that only the complete $2:1$ trion retains quasi-long-range order in the locked phase.
Parameters:
$\rho_c:\rho_b=2:1$,
$U_{bc}=-3$,
$t_{cc}=3$,
$L=160$,
and $t_c=t_b=1$.}
\label{fig:fig3}
\end{figure}

\textit{Correlation Hierarchy.}---To characterize the emergent composite order,
we evaluate the asymptotic spatial correlation functions.
The resulting correlation hierarchy is shown in
Fig.~\ref{fig:fig3}. In the locked phase, the pinning of $\Phi^*_-$ dynamically generates a mass gap $\Delta_{\text{gap}}$. Consequently, its conjugate phase field $\Theta^*_-$ fluctuates wildly, disordering any vertex operator that depends on it. 

Using the re-bosonization phase identity $\theta_P = 2\theta_c$, we express the physical creation operators in terms of the decoupled normal-mode phase fields, $\Theta^*_+$ and $\Theta^*_-$. For single-particle excitations, the operators take the form $c^\dagger \sim e^{-i\theta_P/2}$ and $b^\dagger \sim e^{-i\theta_b}$. When projected onto the normal modes, both operators inherently couple to the disordered relative phase $\Theta^*_-$. As a result, their correlation functions decay exponentially: $G_c(x), G_b(x) \sim \exp(-|x|/\xi)$, where the correlation length is $\xi \sim v^*_- / \Delta_{\text{gap}}$. Similarly, partial composite operators, such as the bare fermion pair ($\Delta_{cc}^\dagger \sim e^{-i\theta_P}$) and the boson-fermion pair ($\Delta_{bc}^\dagger \sim e^{-i(\theta_b + \theta_P/2)}$), also retain a finite dependence on $\Theta^*_-$, leading to rapidly decaying correlations,
as confirmed numerically in
Fig.~\ref{fig:fig3}. Their correlations decay exponentially and therefore do not remain coherent at long distances.

However, the full $2:1$ trionic composite exhibits a simple but precise phase cancellation. Its creation operator is given by $T^\dagger(x) \sim b^\dagger(x) c^\dagger(x) c^\dagger(x+a) \sim e^{-i(\theta_b + \theta_P)}$. Substituting the normal modes into this collective phase yields:
\begin{equation}
\theta_b + \theta_P \propto (\Theta^*_+ - \Theta^*_-) + (\Theta^*_+ + \Theta^*_-) = 2\Theta^*_+.
\end{equation}
The trionic phase completely decouples from the disordered relative mode $\Theta^*_-$, depending exclusively on the gapless acoustic mode $\Theta^*_+$. Consequently, the trionic correlation function uniquely survives the phase fluctuations, exhibiting algebraic decay [Fig.~\ref{fig:fig3}]: 
\begin{equation}
G_{bcc}(x) = \langle T^\dagger(x) T(0) \rangle \sim |x|^{-\eta}.
\end{equation}

This correlation hierarchy shows that only the full trion retains quasi-long-range order in the trionic phase.

\begin{figure}
    \centering
    \includegraphics[width=1\linewidth]{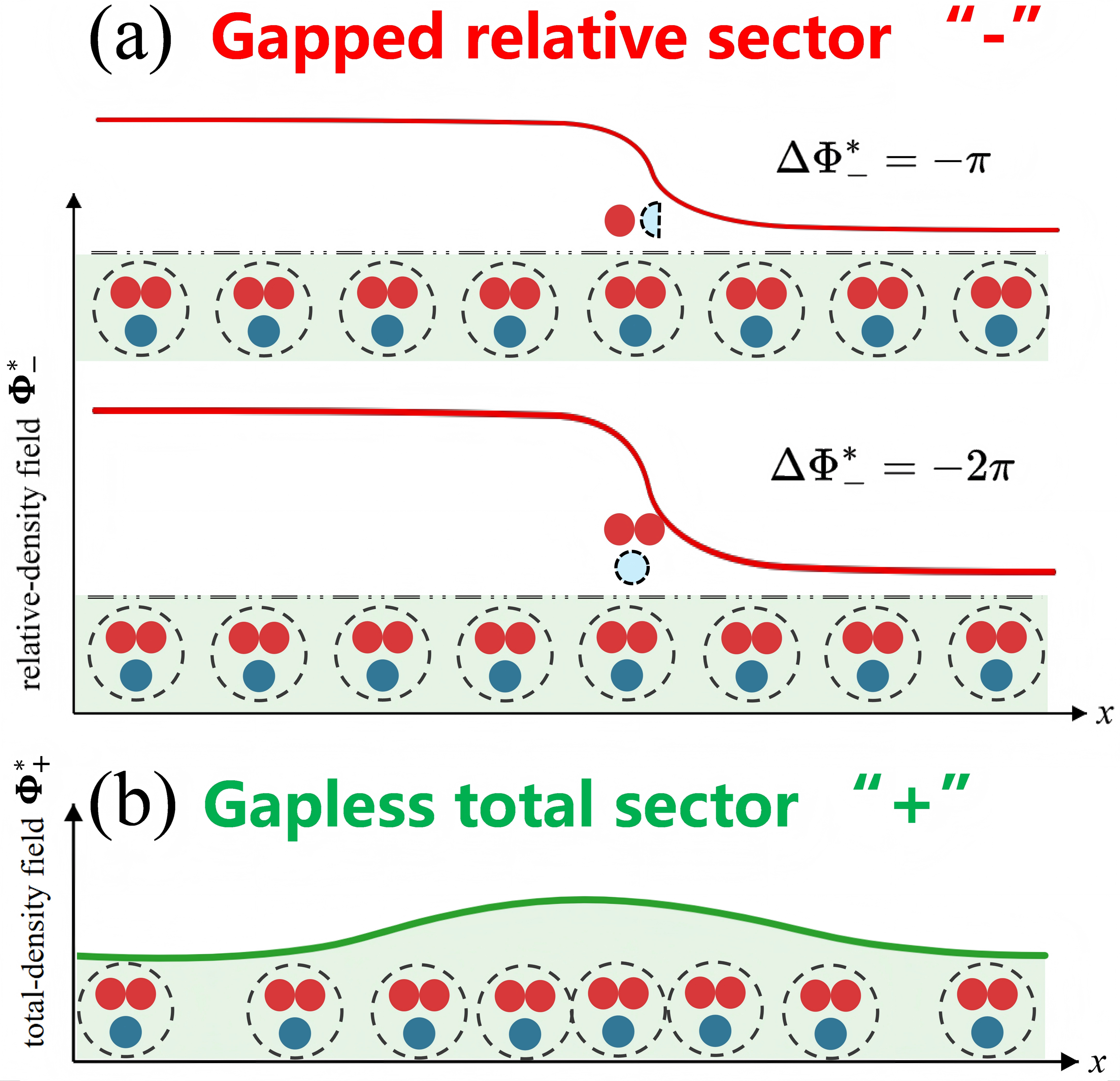}
    \caption{Low-energy excitations of the trionic liquid.
(a) Topological excitations of the gapped relative-density sector $\Phi_-$. The effective locking potential $\cos(2\Phi_-)$ supports domain-wall configurations connecting degenerate vacua. The upper schematic illustrates a candidate geometric $\pi$-kink with formal charge assignment $(\Delta N_c,\Delta N_b)=(1,-1/2)$, while the lower schematic shows the composite $2\pi$ kink carrying $(\Delta N_c,\Delta N_b)=(2,-1)$. Owing to fermion-parity constraints, the $\pi$-kink couples to the massive single-fermion sector, whereas the $2\pi$ kink represents the local topological excitation of the pair-boson liquid.
(b) Gapless total-density sector $\Phi_+$, corresponding to long-wavelength acoustic density fluctuations of the trionic liquid.}
    \label{fig:fig4}
\end{figure}

\textit{Topological Excitations of the trionic Phase.}--- The low-energy excitation spectrum of the trionic liquid is summarized schematically in Fig.~\ref{fig:fig4}. The gapless total-density sector supports acoustic density fluctuations [Fig.~\ref{fig:fig4}(b)], while the gapped relative sector hosts finite-energy topological excitations [Fig.~\ref{fig:fig4}(a)]. These correspond to domain walls (kinks) that spatially interpolate between adjacent degenerate minima of the effective sine-Gordon potential $\cos(2\phi_b - 2\phi_P)$.

A candidate minimal geometric kink [Fig.~\ref{fig:fig4}(a)] corresponds to a phase slip connecting two nearest minima, enforcing the boundary condition $\Delta(\phi_b - \phi_P) = -\pi$. Assuming this localized defect leaves the gapless acoustic normal mode unperturbed ($\Delta \Phi_+^* = 0$, which yields $\Delta \phi_b + \Delta \phi_P = 0$), the individual macroscopic phase shifts are $\Delta \phi_b = -\pi/2$ and $\Delta \phi_P = \pi/2$.

Using the standard bosonization dictionary $\Delta N = \Delta \phi/\pi$, the induced bosonic charge is $\Delta N_b = -1/2$. For the fermionic sector, the infrared density constraint $\phi_P = \phi_c/2$ inherently dictates that the microscopic fermion number shift is twice the pair number shift: $\Delta N_c = 2\Delta N_P$. Substituting $\Delta \phi_P = \pi/2$, we obtain $\Delta N_c = 1$. Thus, the minimal geometric phase slip formally carries the fractionalized charge assignment $(\Delta N_c, \Delta N_b) = (1, -1/2)$.

To determine the physical status of this excitation, we must apply fermion-parity constraints. The formal $\pi$-kink carries an odd fermion parity ($\Delta N_c = 1$). Because the low-energy sector contains only even-parity pair states, an odd-parity kink must involve a broken pair. Therefore, it is not a freely propagating local excitation in the low-energy pair liquid; rather, it couples to the massive single-fermion sector, acting as a domain wall that traps a localized single fermion. The physical local excitation operating entirely within the pair-boson liquid is instead the composite $2\pi$ kink. Carrying the even-parity charge assignment $(2, -1)$, this composite instanton represents the true deconfined topological defect of the phase.


\textit{Experimental Signatures and Conclusion.}---The physical properties of this reconstructed trionic liquid can be probed in state-of-the-art ultracold atomic setups, such as tunable 1D Bose-Fermi mixtures in optical lattices. The correlated pair-hopping $t_{cc}$ can be engineered via periodic lattice driving (Floquet engineering) \cite{Eckardt2017, Gorg2019,Weitenberg2021, Viebahn2021}. Specifically, Quantum gas microscopy \cite{Gross2017, Preiss2015} can reveal the density-profile signatures of the domain walls, while correlation measurements can distinguish the exponential decay of partial channels from the algebraic trionic channel.

In conclusion, we have shown that correlated kinetics provide a powerful mechanism to bypass the geometric obstruction of asymmetric multi-particle locking. By dynamically reorganizing the infrared degrees of freedom, the system maps a bare instability into a stable, effective pair-boson theory. This hierarchical reconstruction stabilizes a robust trionic liquid and dynamically reshapes its physical excitation spectrum, providing a clear condensed-matter example of how hierarchical reconstruction reshapes the excitation spectrum of a composite liquid.


\textit{Acknowledgments.---}We thank Fei Teng for his helpful discussion. This work is supported by the National Key Research and Development Program of China (Grant No. 2022YFA1404204), the National Natural Science Foundation of China (Grant No. 12274086), and the Quantum Science and Technology-National Science and Technology Major Project (Grant No. 2024ZD0300104).

\bibliographystyle{apsrev4-2}
\bibliography{HTFS}

\end{document}